# Expectation-Realization Interpretation of Quantum Superposition

Yanting Wang (王延颋)[1,2,*]

[1]Institute of Theoretical Physics, Chinese Academy of Sciences, Beijing 100190, China

[2]School of Physical Sciences, University of Chinese Academy of Sciences, Beijing 100049, China

**ABSTRACT:** By comparing Schrödinger's cat with its classical counterpart, I show that a quantum superposition should be understood as an expectation over possible eigenstates weighted by wave-like probabilities. Upon the occurrence of a certain event, the quantum system is randomly realized into one of the possible eigenstates due to its intrinsic stochasticity. While the randomness of a single realization cannot be controlled or predicted, the overall distribution can be regulated via experimental setup and converges as the number of events increases. A measurement is indeed an activity employing a certain event to convert a quantum effect into a macroscopic outcome. Consequently, the puzzling concepts of wavefunction collapse, many worlds, and decoherence become unnecessary for understanding quantum superposition. This *expectation-realization interpretation*, which integrates probability theory with wave mechanics, can also be extended to quantum pathways. Moreover, it reframes tests of Bell's inequalities as validating the wave-like probability nature of quantum mechanics, with no need to invoke the mysterious notions of quantum non-locality and "spooky action at a distance".

In 1935, Erwin Schrödinger[1] proposed a famous thought experiment to illustrate the paradoxes of superposition and measurement in quantum mechanics. In this thought experiment, a cat is placed in a gas chamber, where the release of toxic gas, triggered by an atom-decay event, results in the death of the cat. If the quantum state of the atom is denoted as a superposition of the decay state $|C\rangle$ and the non-decay state $|N\rangle$ with the probability amplitudes of $\alpha$ and $\beta$ satisfying the quantum normalization condition

$$|\alpha|^2 + |\beta|^2 = 1, \qquad (1)$$

and the state of the cat is denoted as $|D\rangle$ (dead) and $|A\rangle$ (alive), then the combined quantum state reflecting the entanglement of the atom and cat states is

$$|\psi\rangle = \alpha|C\rangle|D\rangle + \beta|N\rangle|A\rangle. \qquad (2)$$

This appears to lead to a bizarre macroscopic state of the cat, existing in a superposition of both dead and alive. A definite outcome, whether the cat is dead or alive, is only established upon observation when the chamber is opened.

This thought experiment extends the quantum description of microscopic particles to macroscopic objects, and thus leads to the confusion of applying the quantum concepts of superposition and measurement to macroscopic objects which are supposed to obey deterministic classical mechanics. Extensive efforts have been devoted to providing a reasonable interpretation of Schrödinger's cat, including the Copenhagen interpretation[2,3] that the wavefunction collapses to one of its eigenstates upon measurement, the many-worlds interpretation[4-7] that the universe branches into different realities, and the decoherence theory[8-11] that the interaction with the environment destroys the superposition. However, a satisfying interpretation has not yet been formulated, and remains a topic of heated debate alongside other fundamental problems in quantum mechanics[12].

Here I propose a classical version of Schrödinger's cat, illustrated in Fig. 1, to help clarifying several basic concepts necessary for interpreting this thought experiment, and extend them to a deeper understanding of quantum superposition.

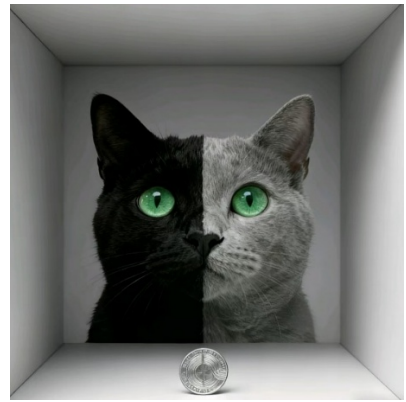

**FIG. 1.** Cartoon illustration of the classical version of Schrödinger's cat. Whether the cat is dead or alive depends on the result of a coin flip.

In contrast to its quantum counterpart, the classical model has the fate of the cat determined by a coin flip in terms of classical probability theory. The cat is killed if the coin lands heads up, denoted by $|C\rangle_c$, and survives if it lands tails up, denoted by $|N\rangle_c$, where the subscript "c" denotes "classical". The probabilities for getting heads and tails are $P_C$ and $P_N$, respectively, satisfying the classical normalization condition

$$P_C + P_N = 1. \quad (3)$$

The flipping of the coin is a specific *event*. Before the event, the *expectation* of the cat state is

$$|\psi\rangle_c = P_C |C\rangle_c |D\rangle_c + P_N |N\rangle_c |A\rangle_c. \quad (4)$$

We can see that, although there are no microscopic particles involved, the macroscopic cat state is still a superposition. After the event occurs (the coin is flipped), the cat's state is randomly *realized* into one of the physical states (classical eigenstates), i.e., dead or alive, with the probability of $P_C$ or $P_N$. Therefore, a superposition state should not be regarded as a physical state, but rather an expectation over possible future outcomes associated with certain probability weights. One of the eigenstates must be realized after an event occurs. Whether the cat is alive or dead is a stochastic consequence of the event and does not depend on the measurement.

The understanding of this *expectation-realization* (ER) *interpretation* may be further simplified by considering a classical die. An ideal die has a uniform distribution of taking one of the discrete values from 1 to 6 with an equal probability of 1/6, so before rolling, the classical superposition of the die can be described as

$$|\psi\rangle_c = \sum_{i=1}^{6} \frac{1}{6} |i\rangle_c, \quad i = 1, 2, \cdots, 6, \quad (5)$$

whose expectation value is 3.5, which cannot really occur in any single rolling event. After an event of rolling, the state of the die must be realized to randomly take one of the six classical eigenstates. The distribution approaches uniformity as the number of rolling events increases.

Until now, I have clarified that the superposition paradox arises from confusing expectation with physical reality. A superposition state is actually an expectation over possible eigenstates before the occurrence of an event, e.g., particle creation/annihilation, collision, scattering, etc. An event drives the system to randomly realize into one of its eigenstates. Both classical probability theory and quantum mechanics share the same above features. The difference between the quantum and classical versions of Schrödinger's cat is that the quantum superposition takes nonlinear wave-like probabilities as Eq. (1), while the classical one is simply a linear superposition as Eq. (3).

Subsequently, the ER interpretation provides a physical picture for quantum superposition as described below. For the nonrelativistic case, a quantum superposition, as can be determined by a given system potential via the Schrödinger equation, is actually the expectation of possible eigenstates. The probability amplitudes obey the laws of wave mechanics (see, e.g., Eq. 1), where coherence comes into play. Upon the occurrence of an event, one of the possible eigenstates is randomly realized. This process is driven by intrinsic stochasticity while remains constrained by certain conservation laws, such as those of energy, momentum, and angular momentum.

From this perspective, quantum mechanics unnecessarily relies on the puzzling concepts of wave-function collapse upon measurement, many copies of real worlds, and decoherence of quantum superposition. As opposed to the Copenhagen interpretation, the ER interpretation regards a quantum system as always residing in one of its eigenstates, rather than collapsing upon measurement. A measurement, in essence, is an activity employing a certain event to convert a quantum effect into a macroscopic outcome. Note that the measurement instrument may alter the Hamiltonian of the entire system, thereby potentially interfering with the original quantum system and leading to different results. Regarding the many-worlds interpretation, it coincides with the ER interpretation if we regard the many worlds as different virtual possibilities which, except the realized one, vanish after an event. As for the decoherence theory, the coherence among different eigenstates, a consequence of wave-like probabilities, does not appear in a single realization, and becomes more significant with increasing number of realizations.

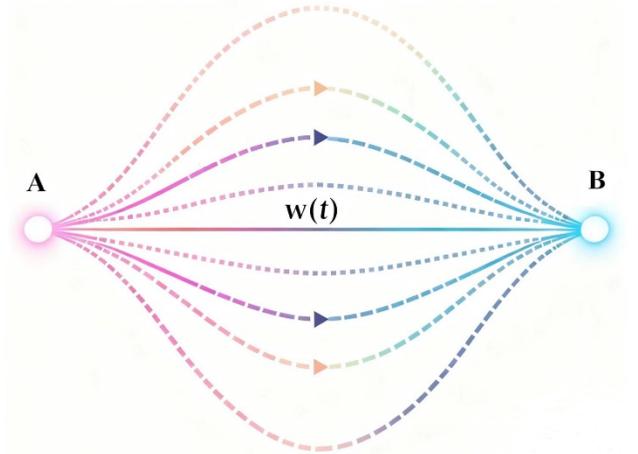

**FIG. 2.** Schematic of quantum pathways going from the initial state A to the final state B. After the occurrence of a certain event, the system is randomly realized into one of the pathways with a certain probability weight $w(t)$.

The ER interpretation can also be extended to understand quantum pathways. As illustrated in Fig. 2, given a set of potentials as well as the initial state (A) and the final state (B) fixed, a quantum system can have enormous number of pathways going from A to B. Governed by the total Hamiltonian, each pathway carries a certain probability weight, which can in principle be quantified via Feynman's path integral method[13], incorporating the coherence between pathways. Each time the system is driven by an event, it stochastically realizes into one of the possible pathways from A to B. The full set of pathways is approached as the number of such realizations increases. For instance, in the double-slit experiment, once a photon is emitted, it stochastically selects one of the possible pathways to pass through one of the two slits before ultimately striking the screen to form a random dot. Interference fringes emerge after a sufficient number of photons have been emitted and passed through the double-slit apparatus. If an additional instrument is introduced to detect which slit a specific photon traverses, the total system Hamiltonian will change, thereby alter the set of possible pathways and their associated probability weights.

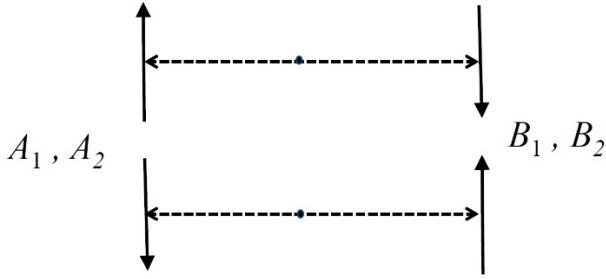

**FIG. 3.** Conceptual diagram of the CHSH setup. Each time a pair of particles with opposite spins is created, the pair occupies one of the two system eigenstates and travels in opposite directions. Alice performs measurements on the left using two distinct schemes, denoted as $A_1$ and $A_2$, while Bob conducts measurements on the right via $B_1$ and $B_2$.

The ER interpretation also provides a way to explain the results of Bell's inequalities[14] without invoking the mysterious notions of quantum nonlocality and "spooky action at a distance". In the Clauser-Horne-Shimony-Holt (CHSH) version of setup[15-21] for testing Bell's inequalities, as shown in Fig. 3, a pair of entangled particles are so created that they are at the singlet state as

$$|\psi\rangle = \frac{1}{\sqrt{2}}(|\uparrow\rangle|\downarrow\rangle - |\downarrow\rangle|\uparrow\rangle), \qquad (6)$$

which means that the two particles always have two opposite spin directions and the system is under the superposition of the two possible spin-pair eigenstates ($|\uparrow\rangle|\downarrow\rangle$ and $|\downarrow\rangle|\uparrow\rangle$). The two particles are then sent out without decoherence in opposite directions, with one detector (Alice) placed on the left and the other (Bob) on the right. The ER interpretation tells us that each time the two particles are created, they are realized into one of the two coherent eigenstates. Note that the entanglement of the two particles is achieved through the system setup of precisely keeping the two-particle system in the superposition state specified in Eq. (6). At the two ends, Alice randomly preforms measurements at two angles represented by the measurement operators $A_1$ and $A_2$, and Bob by $B_1$ and $B_2$. The CHSH operator is defined as

$$S = A_1 \otimes B_1 + A_1 \otimes B_2 + A_2 \otimes B_1 - A_2 \otimes B_2. \qquad (7)$$

Because the eigenvalues of the measurement operators are all $\pm 1$ and $[A_i, B_j] = 0$, $i, j = 1, 2$, in the optimal case, we have[22]

$$S^2 = 4 + [A_1, A_2] \otimes [B_2, B_1]. \qquad (8)$$

Because the norm of the commutators

$$\|[A_1, A_2]\| \leq 2\|A_1\|\|A_2\| = 2, \quad \|[B_2, B_1]\| \leq 2, \qquad (9)$$

we finally have $S^2 \leq 8$, and thus the CHSH value can reach the Tsirelson Bound[23-25] of

$$\|S\| \leq 2\sqrt{2}. \qquad (10)$$

For the classical counterpart, $[A_1, A_2] = [B_2, B_1] = 0$, so the upper bound of $\|S\|$ cannot exceed 2.

Therefore, under the ER interpretation, a single particle pair is already in one of the two eigenstates at the moment of its creation. A measurement does not induce wavefunction collapse but only discloses this pre-existing state. Moreover, the entanglement between the two particles is predetermined before they travel, and is irrelevant to the distance between Alice and Bob as long as no decoherence takes place during the transportation. The fact that the Tsirelson Bound exceeding the classical upper bound is actually attributed to the wave-like probabilities, which leads to the coherence between the two eigenstates of the spin-pair system embedded in Eq. (6) and the local measurement incompatibility at both ends confined by Eq. (9). With the concept of wavefunction collapse being abandoned, the mysterious notions of quantum non-locality and spooky action at a distance are untenable. On the other hand, the hidden variable assumption in the Einstein–Podolsky–Rosen (EPR) paradox[26] is certainly incorrect: The randomness of which eigenstate a single pair occupies cannot be controlled or predicted, and only the overall distribution described by Eq. (6) can be regulated by experimental setup and approached as the number of spin pairs increases.

In summary, by proposing a classical counterpart of the quantum version of Schrödinger's cat, I show that quantum superposition can be well understood by the ER interpretation as arising from the integration of probability theory and wave mechanics. A quantum system always resides in one of its eigenstates, while a superposition state represents a coherent expectation over all possible eigenstates. Upon the occurrence of an event, the quantum system is realized into one of the possible eigenstates. A measurement is an activity employing a certain event to convert a quantum effect into a macroscopic outcome. During a measurement, the instrument may contribute to the system Hamiltonian, so its influence should be accounted for by treating the instrument as part of the overall quantum system. This physical picture also applies to quantum pathways and provides a more natural interpretation of the experimental results for Bell's inequalities.

The author sincerely thanks Prof. Chi Xiong for insightful discussions. This work was financially supported by the National Natural Science Foundation of China (No. 12047503). Figs. 1 and 2 were generated with the assistance of Doubao AI.

———

* Corresponding author: wangyt@itp.ac.cn.